\documentclass[pre,aps,amsmath,reprint,amssymb,showpacs,final]{revtex4-1}
\usepackage{amsfonts}
\usepackage{bm}
\usepackage{dcolumn}
\usepackage{graphicx}
\usepackage{color}
\usepackage{version}
\usepackage{url}
\newcommand{\be}{\begin{equation}}
\newcommand{\ee}{\end{equation}}

\begin{document}

\noindent \textbf{\textit{Accepted for publication in Physical Review E (https://journals.aps.org/pre/})}

\title{Geometrical properties of rigid frictionless granular packings \\ 
as a function of particle size and shape}

\author{Jean-Fran\c cois Camenen}
\affiliation{Universit\'e Bretagne-Sud, IRDL \\
2 rue Coat Saint-Haouen, BP 92116 \\
56321 Lorient Cedex, France}
~~\\
\author{Yannick Descantes}\thanks{Corresponding author.}
\affiliation{LUNAM, IFSTTAR \\
site de Nantes, Route de Bouaye \\
CS4 44344 Bouguenais Cedex, France}\\
~~\\
\date{\today}

\begin{abstract}
Three-dimensional discrete numerical simulation is used to investigate the properties of close-packed frictionless granular assemblies as a function of particle polydispersity and shape. Unlike some experimental results, simulations show that disordered packings of pinacoids (eight-face convex polyhedron) achieve higher solid fraction values than amorphous packings of spherical or rounded particles, thus fulfilling the analogue of Ulam's conjecture stated by Jiao and co-workers for random packings [Y. Jiao and S. Torquato, Phys. Rev. E $\textbf{84}$, $041309$ ($2011$)]. This seeming discrepancy between experimental and numerical results is believed to lie with difficulties in overcoming interparticle friction through experimental densification processes. Moreover, solid fraction is shown to increase further with bidispersity and peak when the volume proportion of small particles reaches $30\%$. Contrarywise, substituting up to $50\%$ of flat pinacoids for isometric ones yields solid fraction decrease, especially when flat particles are also elongated. Nevertheless, particle shape seems to play a minor role on packing solid fraction compared to polydispersity. Additional investigations focused on the packing microstructure confirm that pinacoid packings fulfill the isostatic conjecture and that they are free of order except beyond $30$ to $50\%$ of flat or flat \& elongated polyhedra in the packing. This order increase progressively takes the form of a nematic phase caused by the reorientation of flat or flat \& elongated particles to minimize the packing potential energy. Simultaneously, this reorientation seems to increase the solid fraction value slightly above the maximum achieved by monodisperse isometric pinacoids, as well as the coordination number. Finally, partial substitution of elongated pinacoids for isometric ones has limited effect on packing solid fraction or order.
\end{abstract}


\maketitle

\section{Introduction}

Granular materials cover a large variety of natural and industrial matter as diverse as soil, rock, building materials, cereals, or drug capsules. Controling their density, defined as the volume fraction occupied by grains (also called \emph{solid fraction}), is critical to study their mechanical behavior - which may be gaslike, liquidlike or solidlike - and reduce their manufacturing, storage, or packaging costs. Over the centuries, scientists and engineers have attempted to predict the maximum solid fraction achievable by solidlike assemblies - \emph{ or packings} - of hard particles knowning their individual geometrical characteristics.

Due to their apparent simplicity, monodisperse hard sphere packings have received wide interest and various outstanding solid fraction values have been reported, ranging from $\phi=0.555\pm0.005$ for random loose packings~\cite{onoda_90} up to $\phi=\pi/\sqrt{18}\approx0.74$ for crystal-ordered packings~\cite{hales_05}. In the former case, the packing structure is often stabilised by interparticle friction in a hypostatic state~\cite{makse_00}, that is with less particle constraints than their number of degrees of freedom, whereas the latter case seems to be favored by long term cyclic shear~\cite{Panaitescu_12} and refers to highly hyperstatic structures. In between these two extremes lies the random close packing (RCP) state, equivalently defined as the maximally randomly jammed state~\cite{torquato_00} or stable equilibrium state under isotropic pressure of frictionless particles devoid of crystal nucleus~\cite{roux_04}. Such a state may be repeatedly achieved following experimental protocols, e.g. gently kneading waxed balls enclosed in a rubber membrane~\cite{bernal_60}, pouring particles at a controlled rate into a cylinder~\cite{macrae_61}, or vertically shaking them within a container~\cite{scott_69}, as well as numerically using purely geometrical~\cite{lubachevsky_90} or mechanically-based protocols~\cite{makse_00,silbert_02}. Whatever the protocol, the generally agreed solid fraction of sphere packings in RCP state is $\phi_{RCP}=0.6366\pm0.0005$~\cite{scott_69,cumberland_87,ohern_02,silbert_02} and their structure is known to be isostatic with a mean number of contacts per spheres\textemdash or $coordination$ $number$\textemdash equal to $z=6$.

However, real granular media are very seldom made of spheres. In particular, natural $aggregates$\textemdash i.e., granular materials for construction and civil engineering uses\textemdash may at best be sphere-like (rounded) when extracted from alluvial deposit, but most of them are more or less convex polyhedron-like (angular) as a consequence of their processing from the crushing of massive rock deposits. Unfortunately, literature on assessing the packing solid fraction of convex nonspherical particles is far less abundant. Yet such particles would have higher maximum solid fraction values than spheres according to Ulam's conjecture~\cite{gardner_01}. Indeed, Ulam's conjecture was verified numerically for dense crystal packings of spheroids of axes ratio larger than $\sqrt{3}$~\cite{donev_3_04}, regular tetrahedra dimers~\cite{chen_10}, Platonic and Archimedean solids~\cite{betke_00,degraaf_11}, whose reported maximum solid fraction values are respectively $\phi_{spheroids}\approx0.7707$, $\phi_{tetra}\approx0.8663$ and $\phi_{Pl\&Ar}$ in the range $0.784$ for truncated icosahedra to $1$ for cubes. Ulam's conjecture was also investigated experimentally for rounded frictionless particles such as ellipsoids of axes ratio $1.25:1:0.8$ and angular grains such as tetrahedral dice. Surprisingly, both particle shapes were found able to pack randomly as densely as crystal-ordered spheres, with solid fraction values of $\phi=0.74\pm0.005$ for the former~\cite{man_05}, and an even denser $\phi=0.76\pm0.02$ for the latter~\cite{jaoshvili_10} in the limit of infinite system size. In fact, additional simulations revealed that dense amorphous packings of frictionless non-cubical Platonic or axisymmetric-low-aspect-ratio particles would achieve solid fraction values ranging between those of spheres in the RCP and the dense crystal states~\cite{donev_04,baule_13,jiao_11}. This conclusion led Jiao and co-workers to suggest that, among convex particles with moderate asphericity (low aspect ratio), spheres possess the lowest solid fraction in RCP state, which is known as the analogue of Ulam's conjecture for random packings~\cite{jiao_11}. According to microstructural investigations of numerical packings, the high level of solid fraction observed with nonspherical particles is caused by the higher coordination number needed to constrain their additional rotational degrees of freedom and achieve jamming~\cite{donev_04}. Note however that amorphous packings of frictionless Platonic solids are isostatic~\cite{jaoshvili_10,jiao_11}, whereas axisymmetric particles were found hypostatic at least for aspect ratio smaller than $1.5$~\cite{donev_04,donev_07,baule_13}. 

In order to increase further the representativeness of real granular media in packing solid fraction studies, one shall account for the particle size distribution which is seldom monodisperse, and consider higher particle aspect ratios. Indeed, from field experience with aggregates, polydispersity increases the packing solid fraction~\cite{caquot_37,powers_68} and bidisperse rounded particles tend to pack denser than angular ones~\cite{sedran_94}. More or less empirical models have been designed to mimic these phenomena~\cite{stovall_86,cumberland_87,ouchiyama_89,delarrard_99,liu_02}. A few numerical studies have been published~\cite{roux_07,farr_10}, which confirm higher solid fraction values of bidisperse sphere packings compared to monodisperse ones (up to $0.827$ for a particle diameter ratio of $10$), but little to no similar study dealing with polydisperse polyhedron packing was identified. More literature examines the role played by large particle aspect ratio on packing density. Indeed, frictionless axisymmetric particles with length to diameter ratio larger than $2$ to $3$ were reported to exhibit smaller RCP solid fractions than spheres~\cite{kyrylyuk_11,baule_13} as a consequence of increased excluded volume effects.

For both sphere and convex nonspherical particle packings, numerical approaches proved essential to investigate their microstructure and confirm their amorphous nature. Indeed, the presence of translational order may easily be checked using common statistical tools such as the $pair$ $correlation$ $function$ (or $radial$ $distribution$ $function$), which describes the probability of finding a pair of particles a distance $r$ apart relative to the probability expected for a completely random distribution at the same density~\cite{allen_89}. According to this definition, the pair correlation function of a perfectly amorphous packing is $1$ regardless of $r$ value. The pair correlation function of $dense$ amorphous sphere packings obtained from numerical simulations classically tends towards $1$ beyond a few particle diameters~\cite{silbert_02,agnolin_07}, thus evidencing the absence of long range translational order. Similar results evidencing even less translational order were reported from nonspherical particle packings~\cite{williams_03,jaoshvili_10}, as a consequence of their loss of rotational symmetry. For the latter, further checking is needed to ensure the absence of orientational order, which may be achieved using the $nematic$ $order$ $parameter$ and the $biaxial$ $parameter$~\cite{john_04,camp_97}. These parameters assess the level of alignment of respectively one and two of the particle inertia axes, and they vary between $0$ (no alignment) and $1$ (perfect alignment). Several authors scrutinizing the onset of orientational order in initially random convex nonspherical particle packings subjected to a geometrical densification process have reported amorphous-nematic transition characterized by a nematic order parameter jump in the range $[0.2;0.5]$~\cite{allen_93,camp_97}, whereas the nematic-biaxial transition occured for values of the biaxial parameter in excess of $0.2$~\cite{camp_97}. 

The present paper explores the influence of size, angularity and aspect ratio of frictionless particles on their ability to achieve maximally randomly jammed packings. Further to determining maximum solid fraction values, our goal is to understand how the packing microstructure is affected by particle geometrical characteristics. For this purpose, numerical simulations have been performed so as to mimic a mechanically-based densification process, thus allowing comparisons with similar laboratory experiences. In section~\ref{sec:simmet}, we describe our simulation protocol. Results gathered in section~\ref{sec:results} are presented and discussed according to the following outline: in section~\ref{sec:mech_equil}, the mechanical equilibrium achieved by the packings are carefully examined, as well as their homogeneity to ensure the absence of particle segregation; then section~\ref{sec:solid_fraction} reports on the calculated solid fraction values and their variations as a function of particle bidispersity, angularity and aspect ratio; finally, section~\ref{sec:packing_micro} investigates the microstructure of our packings.


\section{Simulation protocol}\label{sec:simmet}

\begin{figure}[!t]
	\centering
\begin{tabular}{c}
\includegraphics*[width=0.65\columnwidth]{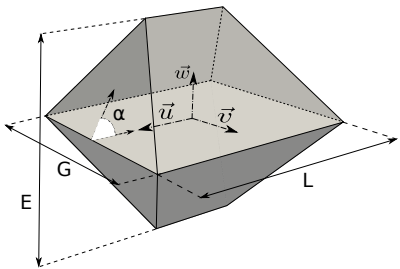}\\
\end{tabular}
\caption{\label{pinacoid}Pinacoid, a model polyhedron characterised by its length $L$, width $G$, height $E$ and angle $\alpha$. This polyhedron has three symmetry planes, each perpendicular to an inertia axis $\roarrow u$, $\roarrow v$, or $\roarrow w$.} 
\end{figure}

The simulated systems are dense assemblies of $3000$ rigid frictionless particles of identical mass density $\rho$, interacting with each other through totally inelastic collisions. Particles are either spheres or $pinacoids$, the latter referring to a variety of convex polyhedra comprised of eight vertices, fourteen edges and eight faces as shown on figure~\ref{pinacoid}. According to an extensive experimental study with various rock types mentioned in ref.~\cite{tourenq_82}, the pinacoid gives the best fit among simple geometries for an aggregate grain. As explained elsewhere~\cite{azema_12,camenen_12}, a pinacoid has three planes of symmetry and is determined by four parameters, length $L$, width $G$, height $E$ ($L \geq G \geq E$) and angle $\alpha$ set to $60^\circ$ here, so that its volume $V$ is given by: 
~~\\
\be V=\frac{EG}{2}(L-\frac{E}{3\sqrt{3}}). \ee \label{equ_1}
~~\\

\begin{figure}[!t]
	\centering
\begin{tabular}{c c}
\includegraphics*[width=0.3\columnwidth]{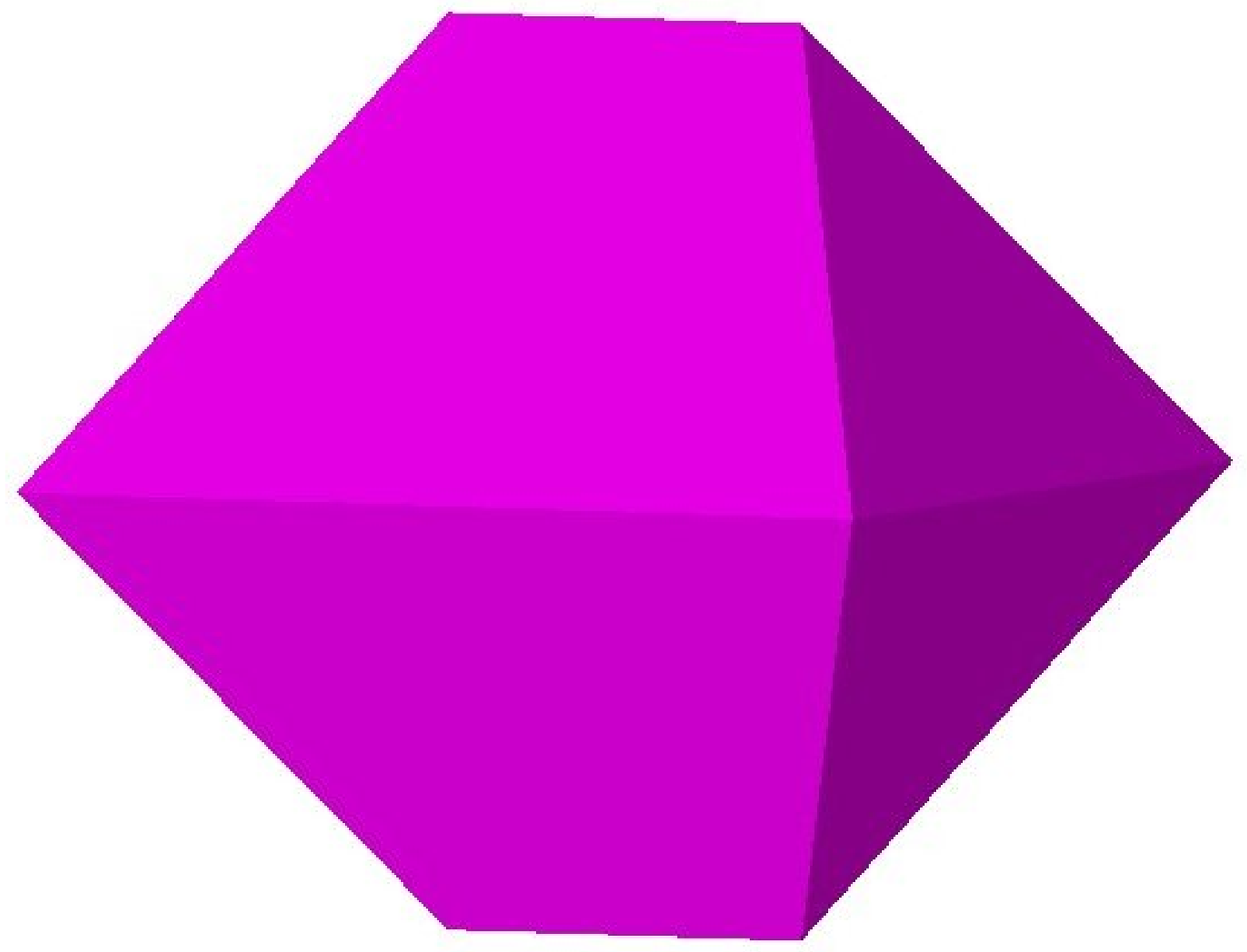}&\includegraphics*[width=0.5\columnwidth]{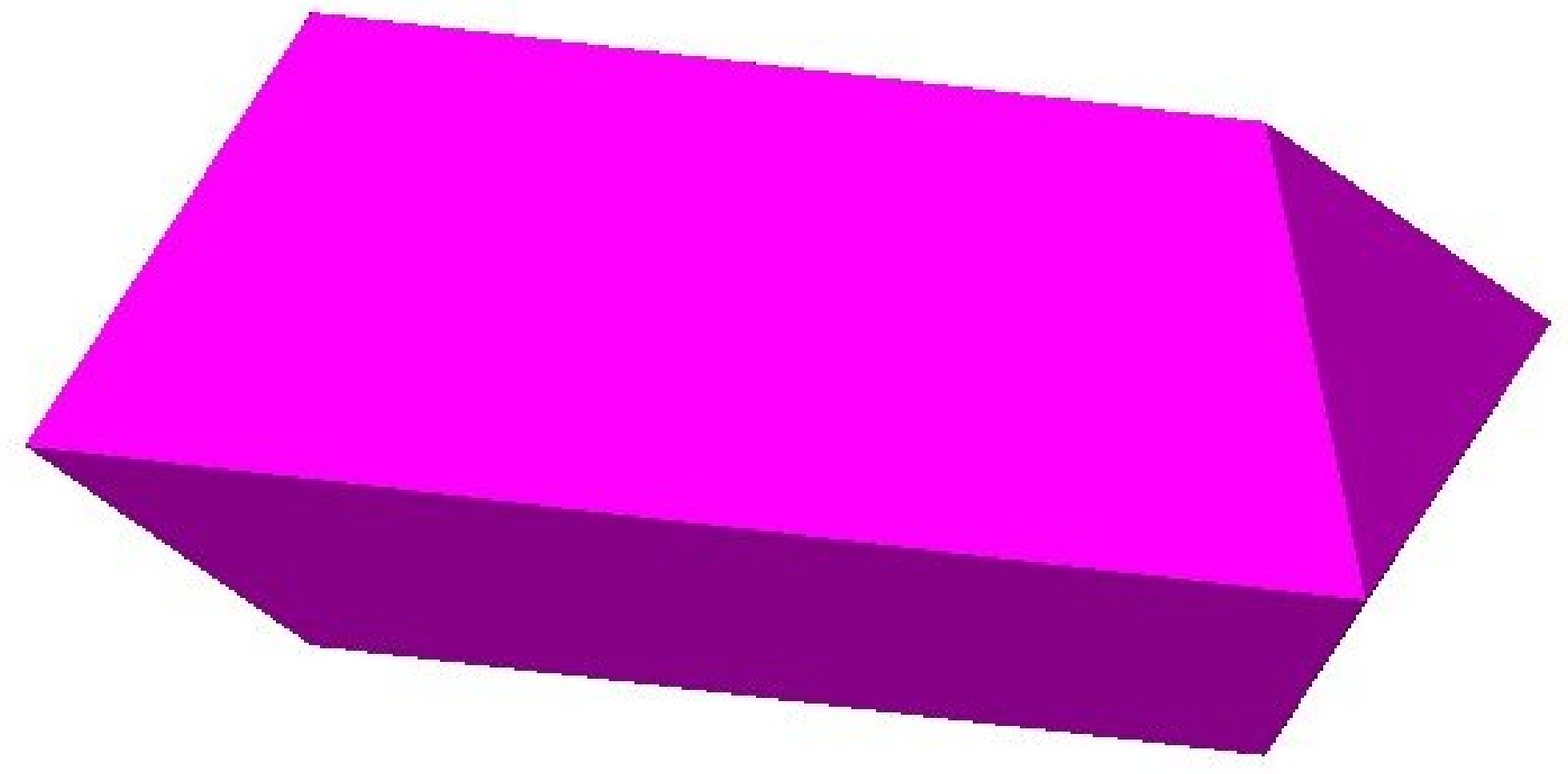}\\
(a) & (b)\\
\includegraphics*[width=0.3\columnwidth]{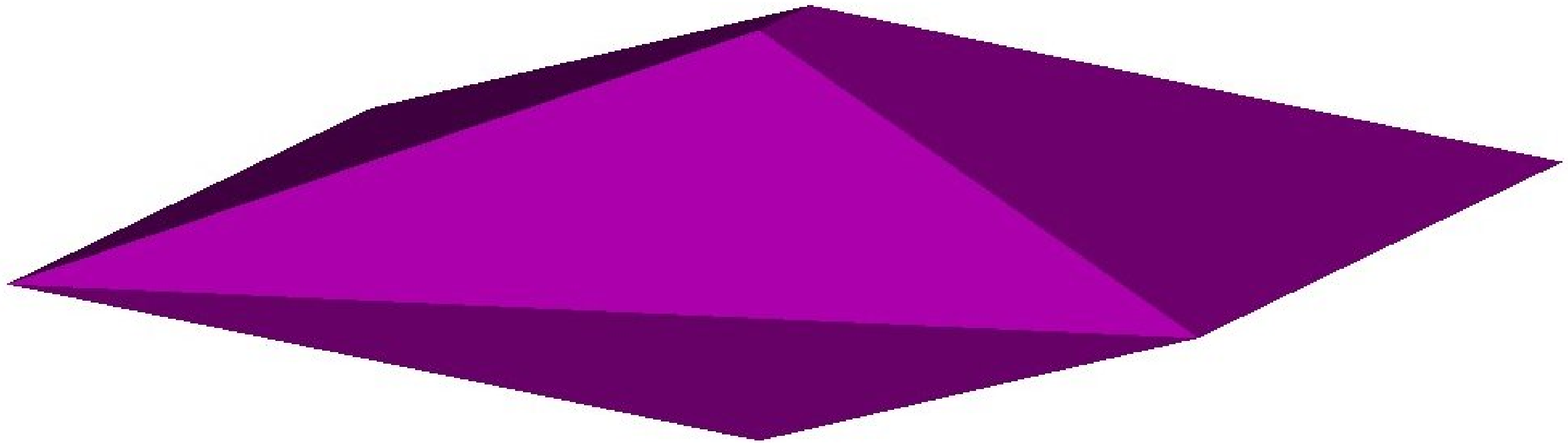}&\includegraphics*[width=0.5\columnwidth]{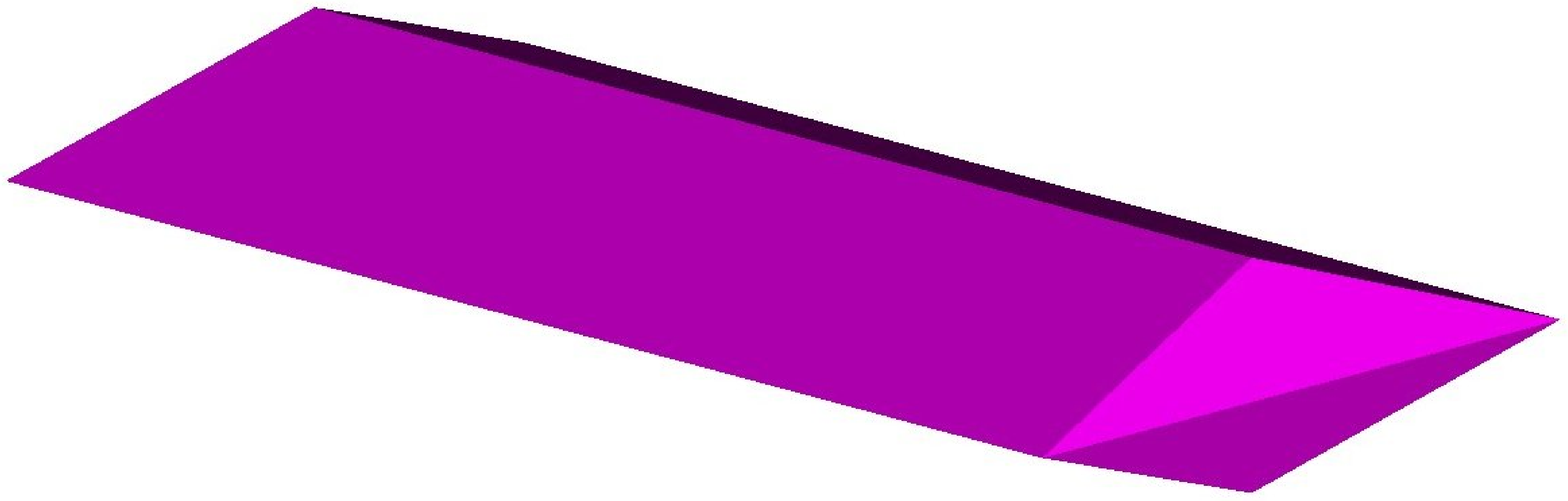}\\
(c) & (d)
\end{tabular}
\caption{\label{fig:pinacoids}(Color online) Snapshot of the pinacoids used in this study: $(a)$ isometric, $(b)$ elongated, $(c)$ flat and $(d)$ flat \& elongated.} 
\end{figure}

\noindent Alternatively, a pinacoid may be determined by its size $d=\sqrt{G^2+L^2}$, corresponding to the diameter of its circumscribed sphere, supplemented with its aspect ratios $L/G$ and $G/E$. Figure~\ref{fig:pinacoids} depicts pinacoids with various aspect ratios and Table~\ref{tab:shape} summarizes the size and aspect ratio of any particle used in the present study. 

\begin{table}[!t]
\begin{center}
        \caption{\label{tab:shape}Size (diameter $d$ for spheres, $d=\sqrt{L^2+G^2}$ for pinacoids) and aspect ratio of simulated particles.}
  \begin{tabular}{lccc}
\hline
\hline
      Particles & Size & $L/G$ & $G/E$\\
\hline

spheres (large, small) & $d$, $d/3$ & $-$ & $-$ \\

isometric pinacoids (large, small) & $d$, $d/3$ & $1$ & $1$ \\

elongated pinacoids & $d$ & $2$ & $1$ \\

flat pinacoids & $d$ &$1$ & $3$  \\

flat \& elongated pinacoids & $d$ & $2$ & $3$ \\
\hline
\hline
  \end{tabular}
\end{center}
\end{table}

Binary mixes in which various proportions of small spheres have been substituted for large ones, or small isometric pinacoids for large isometric pinacoids, have been prepared to study the influence of particle size distribution and angularity on packing properties. Additionally, binary mixes comprising various proportions of flat, elongated, or flat \& elongated pinacoids substituted for large isometric ones have been prepared to investigate the role played by particle aspect ratio. Table~\ref{tab:granulo} summarizes the mean proportion by volume of small, elongated, flat or flat \& elongated particles in each binary mix simulated.

\begin{table}[!t]
\caption{\label{tab:granulo}Mean proportion by volume of small $X_{d/3}$, elongated $X_{P}$, flat $X_{O}$, or flat \& elongated $X_{PO}$ particles substituted for large isometric ones in each binary mix simulated (e.g. $X_P=14\%$ refers to the bidisperse pinacoid packing comprising $14\%$ of elongated pinacoids and $86\%$ of large isometric ones).}
  \begin{tabular}{c|c|c|c|c}
\hline
\hline
Spheres (\%)&\multicolumn{4}{c}{Pinacoids (\%)} \\
\hline
    $X_{d/3} $ & $X_{d/3}$ & $X_{P}$ & $X_{O}$ & $X_{PO}$ \\
\hline
  $0$  & $0$  & $14$  & $12$  & $14$  \\

  $3$  & $13$ & $20$  & $28$  & $29$  \\

  $13$ & $30$ & $36$  & $47$  & $50$  \\

  $23$ & $51$ & $49$  & $77$  & $70$  \\

  $30$ & $64$ & $70$  & $92$  & $100$ \\

  $41$ &  -   & $100$ & $100$ &  -    \\

  $55$ &  -   & -     & -     &  -    \\
\hline
\hline
  \end{tabular}
\end{table}

\begin{figure}[!t]
	\centering
\includegraphics[width=0.7\columnwidth]{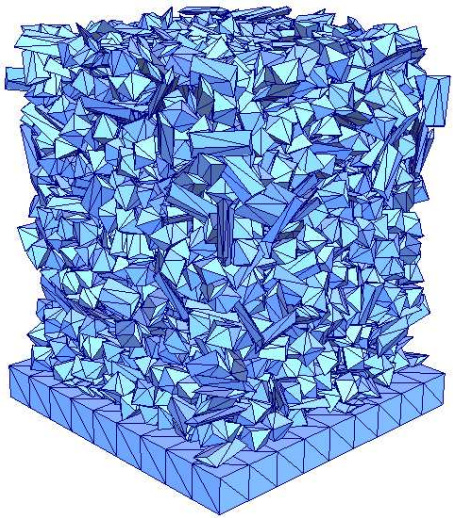}\\
\caption{(Color online) 3D snapshot of a packing incorporating $29\%$ by volume of flat \& elongated pinacoids. Bi-periodic boundary conditions apply in $x$ and $y$ directions}\label{fig:rendus1}
\end{figure}

For each binary mix, three cuboidal samples have been prepared following a pluviation protocol inspired by~\cite{laniel_07} and described in details in~\cite{azema_12,camenen_12}: spherical shells, each circumscribed to a randomly oriented particle, are first randomly dropped inside a vertical parallelepiped container and subsequently moved to the closest local minimum of potential energy; second, the spherical shells are removed, bi-periodic boundary conditions are substituted for the container vertical walls, and the same gravity $\vec g$ ($0$,$0$,$-g$) is applied to all particles until they find an equilibrium position under their own weight (examples of such an equilibrated packing is shown on Fig.~\ref{fig:rendus1}). This protocol was selected for the ability of pluviation to achieve $RCP$ states of hard spheres with solid fraction repeatedly peaking at 
$0.64$~\cite{bernal_60,macrae_61,scott_69,zhang_01,silbert_02,emam_05}. Similarly, nonspherical frictionless particles subjected to the selected pluviation protocol were expected to achieve repeatedly states of maximum density characterized by a unique solid fraction value, provided that they are homogeneously spread and achieve a stable equilibrium without crystallization or segregation~\cite{roux_04,agnolin_07}.

All simulations were performed using the Non Smooth Contact Dynamic method (NSCD)~\cite{radjai_11,radjai_09,jean_99,moreau_94}. This distinct element method (DEM), implemented in the LMGC90 sofware platform~\cite{dubois_03}, was successfully applied to a number of physical problems ranging from dense inertial flows~\cite{lois_05,chevoir_01,azema_12} to quasistatic deformable packings~\cite{azema_07b,azema_09,estrada_08,camenen_12}. Basically, the equations of motion of a collection of rigid particles interacting through unilateral contacts with dry friction are first integrated over one time step. Hence, instead of accelerations and forces, the unknowns are particle velocities and percussions (also called $impulses$). The advantage of this integral formulation is that collision (shock) and lasting contact situations do not need to be distinguished any more, both giving rise to a percussion. Percussions, which may be defined as the integral of a force over one time step, are parameterized using normal ($e_N$) and tangential ($e_T$) restitution coefficients as well as a sliding friction coefficient ($\mu$). Second, at each time step, a geometrical contact detection algorithm identifies any potential contact situation\textemdash mainly vertex-face, edge-edge, edge-face or face-face contacts between convex polyhedra and a point between spheres\textemdash and describes it in terms of location and normal unit vector. Last, the equations of motion are solved at each time step by an iterative process using a non-linear Gauss-Seidel like method. This resolution is performed with respect to $complementarity$ $relations$~\cite{moreau_88} between relative velocities and percussions substituted for classical non-interpenetration (Signorini) and contact friction (Coulomb) constraints.

Simulation parameters are summarized in table~\ref{tab:summary}.

\begin{table}[!t]
  \caption{ \label{tab:summary}Summary of simulation parameters. $3000$ particles, horizontal dimension of square simulation cell $L$ normalized by particle size $d$ (periodic boundary conditions apply along $x$ and $y$ directions), friction $\mu$, restitution coefficients $e_{N,T}$ and time steps $\Delta T$ normalized by $\sqrt{d/g}$.}
  \begin{tabular}{lcccc}
    \hline
    \hline
    Particle            & $L/d$         & $\mu$ & $e_{N,T}$ & $\Delta T/\sqrt{d/g}$ \\
    \hline
    Spheres and         & $3$ to $8$    & $0$   & $0$       & $5\times10^{-3}$      \\
		isometric pinacoids &               &       &           &                       \\
    elongated pinacoids   & $8$ to $11$   & $0$   & $0$       & $5\times10^{-3}$      \\
    flat pinacoids    & $6$ to $8$    & $0$   & $0$       & $5\times10^{-3}$      \\
    flat \& elongated   & $8$           & $0$   & $0$       & $5\times10^{-3}$      \\
		pinacoids           &               &       &           &                       \\
    \hline
    \hline
    \end{tabular}
\end{table}

\section{Results}\label{sec:results}

We first check the quality of mechanical equilibrium achieved by our simulated packings, then we examine the variations of packing solid fraction as a function of particle size and shape. Finally, we investigate microstructural properties of our pinacoid packings to help understanding solid fraction variations.

\subsection{Mechanical equilibrium and homogeneity}\label{sec:mech_equil}
 
A packing of rigid particles achieves a mechanically stable equilibrium under the following conditions: $1$) the net force and net torque applied to each particle as well as its kinetic energy are negligible, and $2$) the potential energy of the collection of particles is minimal. For rigid spheres, all these conditions are met when the following criteria are fulfilled~\cite{camenen_12,agnolin_07}:
\begin{eqnarray}
\sum F  < 10^{-4}d^2P, \label{equ_2} \\
\sum M < 10^{-4}d^3P, \label{equ_3} \\
E_c < 10^{-8}d^3P, \label{equ_4} \\
\sum\limits_{particles} \rho V g = F_S. \label{equ_5}
\end{eqnarray}

\noindent where $\sum F$, $\sum M$ and $E_c$ are respectively the net force, net torque and total kinetic energy of a particle, $P$ denotes the local average stress in its neighborhood and $F_S$ the net force exerted at contacts between particles and the bottom wall. Though our packings of frictionless spheres were found to meet these criteria, moderate deviations were observed for pinacoids. Indeed, the net force was found in the order of $10^{-3}d^2P$, whereas the net torque and kinetic energy were found in the order of $10^{-3}d^3P$ and $10^{-7}d^3P$ respectively. The fact that these results are slightly higher than previous ones~\cite{camenen_12} may be explained by shorter simulation durations (about $25 s$ compared to about $35 s$ in~\cite{camenen_12}), but this does not call into question the quality of mechanical equilibrium achieved (this was further checked upon continuing four simulations up to $35s$ - results are not shown here). Besides, all packings meet the requirement stated by equation~(\ref{equ_5}) since the ratio of the total mass of particles to the net force $F_S$ exerted on the bottom wall varies between $0.9957$ and $1.0048$.

Next, we verify that no significant interpenetration occurs at interparticle contacts to ensure the relevance of solid fraction calculations.  For sphere packings, the maximum calculated interpenetration is lower than $10^{-3}d$. Similarly, we verify that the interpenetration routinely generated at polyhedra contacts when applying the NSCD method~\cite{saussine_04,camenen_12} remains small enough. For this purpose, we apply the virtual works principle to generate a vertical expansion of each packing that will reset the vertical component of all interpenetrations to zero:
~~\\
\be k\sum\limits_{particles} \rho V g z = \sum\limits_{contacts} F_C^z \delta_C^z \label{virt_work}\ee 
~~\\
\noindent where $k$ is the coefficient of vertical expansion of the packing (taken proportional to distance $z$ between the center of gravity of each particle and the bottom wall), $F_C^z$ and $\delta_C^z$ are the vertical components of contact forces and interpenetrations respectively, and both sums are calculated two particle size above the layered bottom to get rid of wall effect~\cite{camenen_12,donev_04}. The mean $k$ value calculated over all pinacoid packings is $0.80\%$ and the sandard deviation is $0.47\%$. Upon assuming a similar coefficient value along $x$ and $y$ axes, interpenetrations at contacts are found to yield no more than $2$ to $3\%$ overestimation in solid fraction calculations, which is consistent with references~\cite{saussine_04,camenen_12}.

\begin{figure}[!t]
	\centering
\includegraphics*[width=0.9\columnwidth]{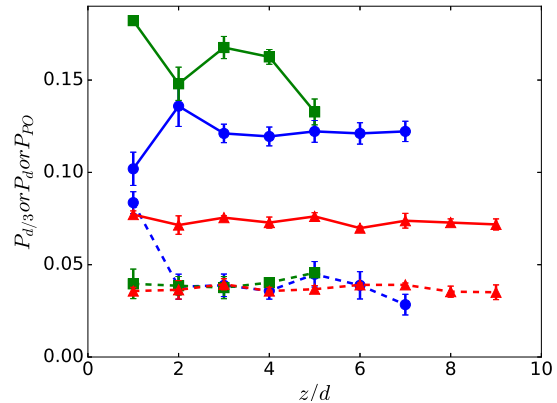}\\
\caption{\label{fig:homo_distri}(Color online) Proportions of small (solid lines) and large (dashed lines) particles in $d$-thick layers along the $z$ axis for bidisperse packings incorporating $X_{d/3}=13\%$ by mass of spheres (${\color{blue}\bullet}$) or pinacoids (${\color[rgb]{0,.5,0}{\blacksquare}}$). Proportions of large isometric pinacoids (solid line) and flat \& elongated pinacoids (dashed line) in $d$-thick layers for packings incorporating $X_{PO}=29\%$ by mass of flat \& elongated (${\color{red}\blacktriangle}$) particles. Error bars denote the standard deviation.}
\end{figure}

Finally, we checked the homogeneity of the distribution of each population of particles once our packings have reached a stable mechanical equilibrium. Figure~\ref{fig:homo_distri} depicts the proportions of large $P_d$, small $P_{d/3}$ and flat \& elongated $P_{PO}$ particles as a function of distance $z$ from the bottom wall of their respective packings. The proportion profiles are almost constant, with only small deviations close to the bottom wall of bidisperse packings. Hence, we can conclude that each population of spheres or pinacoids is reasonably homogeneously spread in its respective packing.

\subsection{Solid fraction}\label{sec:solid_fraction}

In this subsection, we investigate the effect of particle size distribution, angularity, and aspect ratio on the maximum solid fraction value reached by the packing once a stable mechanical equilibrium is achieved. For each packing, the solid fraction is computed upon evenly slicing the packing horizontally and performing analytical calculation of the volume of each sphere or each pinacoid present in each slice of known volume. Naturally, the mean solid fraction is calculated in the bulk, that is away from bottom wall and free surface.

\subsubsection{Effects of particle size distribution and angularity}\label{subsec:ang_poly}

\begin{figure}[!t]
	\centering
\begin{tabular}{c}
\includegraphics*[width=0.9\columnwidth]{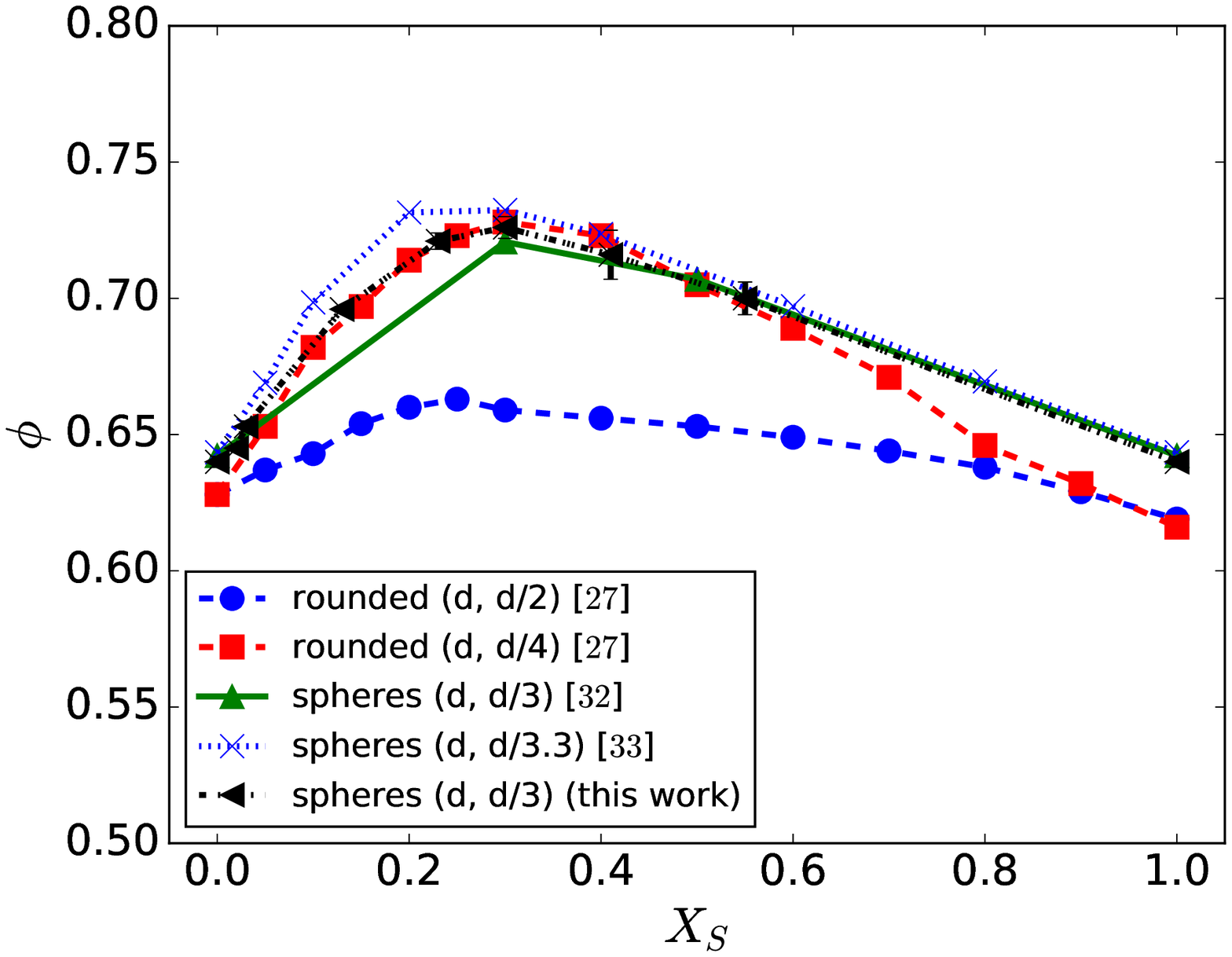}\\
(a) \\
\includegraphics*[width=0.9\columnwidth]{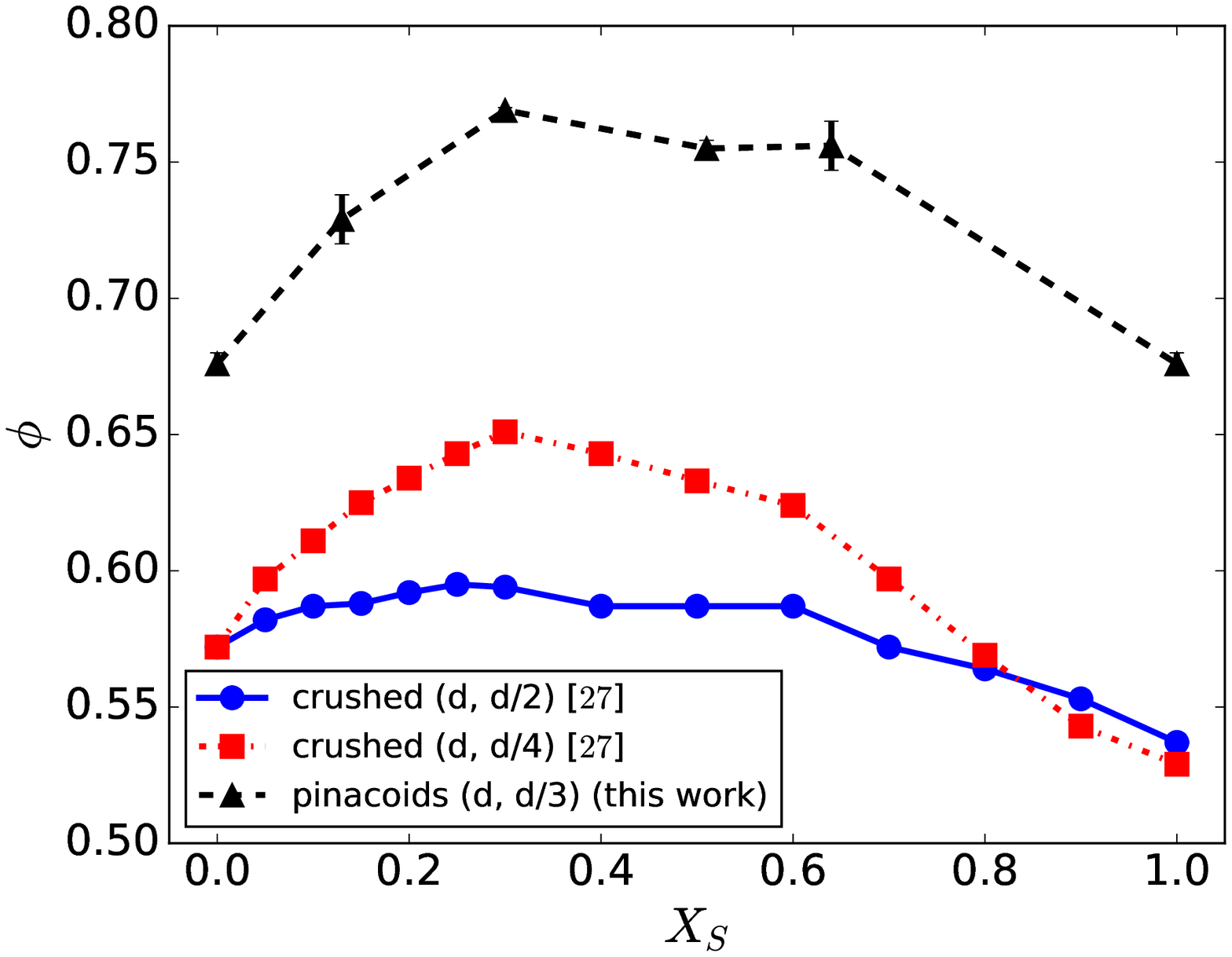}\\
(b) \\
\end{tabular}
\caption{ \label{fig:phi_particles}(Color online) Maximum solid fraction $\phi$ as a function of the volume proportion $X_S$ of (a) small spheres or rounded particles and (b) small pinacoids or crushed particles in bidisperse packings. Error bars denote the standard deviation.}
\end{figure}

For various proportions of small particles $X_S$, Fig.~\ref{fig:phi_particles} depicts maximal solid fraction values $\phi$ achieved by bidisperse packings of (a) spherical particles and (b) pinacoids. For spherical particles, excellent agreement can be observed between our results and those calculated from molecular dynamics~\cite{roux_07} or sphere mapping~\cite{farr_10} simulations. The corresponding curves peak for $30\%$ of small spherical particles by volume, and their respective maximum values are $\phi = 0.726\pm 0.004$ for the present study, $\phi = 0.7207\pm 0.0004$ for ref.~\cite{roux_07} and $\phi = 0.7324$ for ref.~\cite{farr_10}. A good agreement is also observed with experimental results reported in ref.~\cite{sedran_94} from bidisperse packings of ($d$, $d/4$) rounded aggregates densified inside a rigid cylinder through vertical taps and upper surface loading, though these are finite size packings of non-strictly bidisperse spherical particles. By contrast, similar experiments with bidisperse packings of ($d$, $d/2$) rounded aggregates yielded significantly lower maximum solid fractions, which may be explained by increased crowding of the local arrangement of large particles by small ones when their size ratio comes closer to $1$~\cite{stovall_86}. 

For pinacoids, similar layout of the solid fraction curve is observed (Fig.~\ref{fig:phi_particles}b), with monodisperse isometric pinacoid packings ($X_S=0\%$ and $X_S=100\%$) achieving the minimum $\phi = 0.676\pm0.004$ and a peak $\phi=0.769\pm0.001$ being reached when $X_S=30\%$. Interestingly, observe that this peak is significantly higher than the one calculated for sphere packings, suggesting that angular particles pack denser than spherical ones. This conclusion is consistent with previous experimentations reported by ref.~\cite{jaoshvili_10}, showing that monodisperse plastic tetrahedra pack with a maximum solid fraction of $\phi=0.76\pm0.02$ by extrapolation to infinite packing size, and with subsequent numerical work on Platonic solids reported in ref.~\cite{jiao_11}. Note however that other previous works support experimentally~\cite{baker_10} or 
numerically~\cite{smith_10} the opposite conclusion illustrated on Fig.~\ref{fig:phi_particles} by experimental results from ref.~\cite{sedran_94}, according to which rounded particles would pack denser than angular ones. In fact, low experimental solid fraction values result from difficulties in overcoming interparticle friction while compacting a granular assembly~\cite{silbert_02,agnolin_07}, which do not affect our frictionless pinacoids.

\subsubsection{Shape effect}

\begin{figure}[!t]
	\centering
  \includegraphics*[width=0.9\columnwidth]{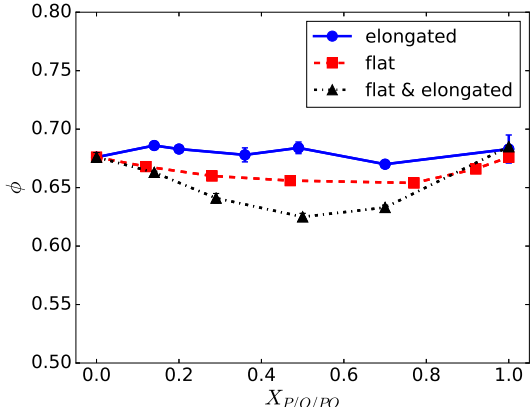}\\
  \caption{\label{fig:phi_L_P_PO}(Color online) Maximum solid fraction $\phi$ $\emph{vs}$ the proportion by volume of (a) elongated $X_P$, (b) flat $X_O$, and (c) flat \& elongated $X_{PO}$ particles in mixtures with large isometric pinacoids. Error bars denote the standard deviation.}
\end{figure}

Figure~\ref{fig:phi_L_P_PO} depicts maximum solid fraction $\phi$ versus the proportion by volume of $(a)$ elongated, $(b)$ flat and $(c)$ flat \& elongated particles in mixtures with large isometric pinacoids. This figure shows that partial substitution of flat pinacoids for isometric ones results in packing solid fraction decrease by maximum $3\%$ for elongated and $8\%$ for flat \& elongated particles. In both cases, the minimum solid fraction is achieved when the proportion of flat or flat \& elongated particles reaches approximately $50\%$. Increasing further the proportion of flat particles raises the solid fraction value up to $\phi = 0.676\pm0.001$ for $100\%$ flat pinacoids and slightly above for $100\%$ flat \& elongated pinacoids. By comparison, random jammed assemblies of $100\%$ oblate ellipsoids with an aspect ratio of $3$ were reported in ref.~\cite{donev_04} to pack with a maximum solid fraction $\phi = 0.67$. Contrarywise, partial substitution of elongated pinacoids for isometric ones does not seem to significantly affect the packing solid fraction which may be estimated at $\phi = 0.683\pm0.012$ for $100\%$ elongated pinacoids. Note that this estimate compares well with the $\phi = 0.68$ solid fraction reported in ref.~\cite{kyrylyuk_11} for packings of $100\%$ spherocylinders with the same length to diameter aspect ratio of $2$. Upon applying the virtual works principle as depicted in equation (\ref{virt_work}), lower bounds accounting for particle maximum interpenetration may be calculated for the solid fraction of packings made of $100\%$ of the various particle shapes studied. Table~\ref{tab:sol_fract} gathers these lower bounds and evidences that, regardless of their aspect ratios, pinacoids pack denser than spheres in the RCP state.

\begin{table}[!t]
  \caption{ \label{tab:sol_fract}Minimum solid fraction values obtained for packings made of $100\%$ isometric, $100\%$ elongated, $100\%$ flat or $100\%$ flat \& elongated pinacoids.}
  \begin{tabular}{l|c|c|c|c}
    \hline
    \hline
    Pinacoid shape   &  $100\%$  &  $100\%$ &  $100\%$  &  $100\%$  \\
                                    & isometric  &  elongated &  flat  & flat \& elongated  \\
    \hline
    $\phi_{min}$    & $0.664$  & $0.662$   & $0.648$       & $0.654$      \\
    \hline
    \hline
    \end{tabular}
\end{table}

\subsection{Microstructure}\label{sec:packing_micro}

To shed some light on maximum solid fraction variations as a function of particle size, angularity and shape, we now investigate the microstructural properties of our packings. We first examine the existence of particle arrangement and then we explore the contact network.


\subsubsection{Particle arrangement}

\begin{figure}[!t]
	\centering
\includegraphics*[width=0.9\columnwidth]{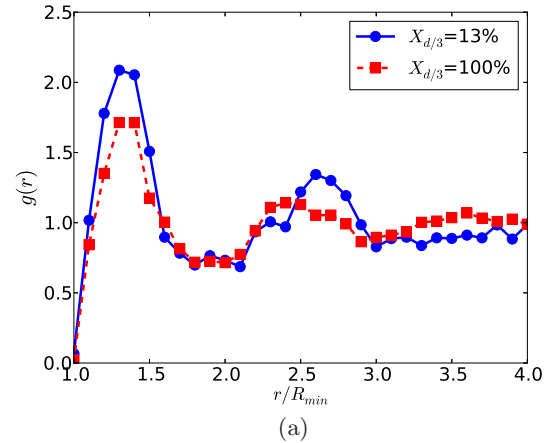}\\
(a) \\
\includegraphics*[width=0.9\columnwidth]{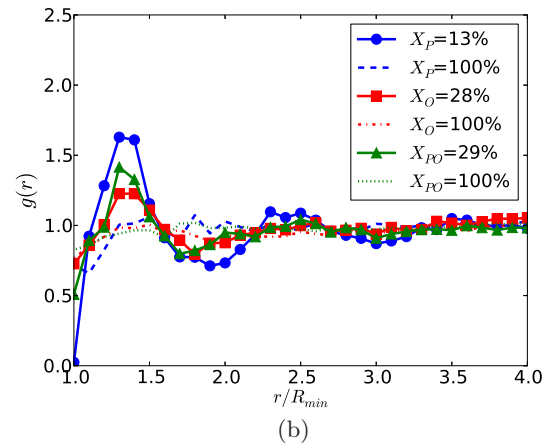}\\
(b) \\
\caption{\label{fig:gr}[Color online) Pair correlation function $g(r)$ for several proportions by volume of $(a)$ small pinacoids $X_S$ and $(b)$ elongated $X_P$, flat $X_O$ or flat \& elongated particles $X_{PO}$. For each mixture, $R_{min}$ stands for
the radius of the sphere circumscribed to the smallest pinacoid (e.g. $R_{min}=d/6$ for mixtures of large and small isometric pinacoids).}
\end{figure}

\begin{figure}[!t]
	\centering
\includegraphics*[width=0.5\columnwidth]{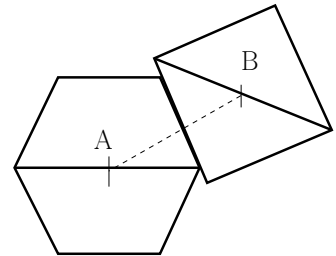}\\
\caption{\label{fig:gr_face_face}Cross-sectionnal view of the arrangement due to contact between trapezoidal and triangular faces with a distance between pinacoid centers of $r \simeq (2/\sqrt6 + 1/2)R_{min} \simeq 1.3R_{min}$.}
\end{figure}

The existence of translational arrangement is studied by means of the pair correlation function $g(r)$, which is calculated in the packing bulk (two particle size above the layered bottom to get rid of wall effect~\cite{camenen_12,donev_04}) using the expression detailed in ref.~\cite{allen_89} (page $55$). Figure~\ref{fig:gr} depicts the variations of $g(r)$ in (a) bidisperse pinacoid packings and in (b) mixtures of isometric pinacoids with elongated, flat or flat \& elongated particles. These curves clearly show that, for $r$ larger that $2R_{min}$ to $3R_{min}$, $g(r)$ no more significantly differs from $1$, meaning that the local distribution of particle centers is free of long-range translational ordering. Like tetrahedron packings~\cite{jaoshvili_10,smith_10,neudecker_13}, pinacoid packings appear to be less correlated than sphere packings, because the former lack of rotational symmetry and angles between their adjacent flat faces induce frustration. Note that the peak visible on several curves for $r \simeq (2/\sqrt6 + 1/2)R_{min} \simeq 1.3R_{min}$ corresponds to the face-face contact between small particles as depicted on fig.~\ref{fig:gr_face_face}.

\begin{figure}[!t]
	\centering
\includegraphics*[width=0.9\columnwidth]{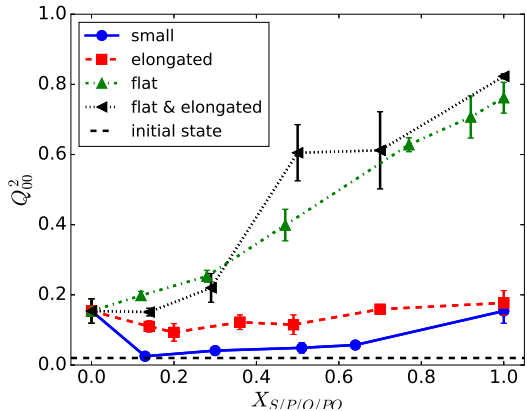}
\caption{\label{fig:Qoo}(Color online) Nematic order parameter $(Q_{00}^2)$ of pinacoid packings incorporating various proportions by volume of small ($X_S$), elongated ($X_P$), flat ($X_O$) or flat \& elongated ($X_{PO}$) pinacoids. Error bars denote the standard deviation. Initial state refers to $Q_{00}^2$ values calculated during sample preparation just before applying gravity.}
\end{figure}

\begin{figure}[!t]
	\centering
\includegraphics[width=0.7\columnwidth]{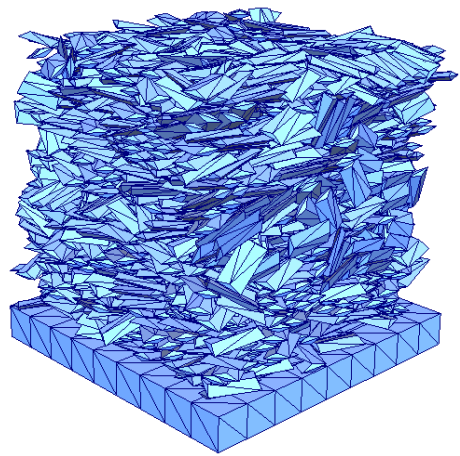}\\
\caption{(Color online) 3D snapshot of a packing incorporating $100\%$ by volume of flat \& elongated pinacoids. Bi-periodic boundary conditions apply in $x$ and $y$ directions}\label{fig:rendus2}
\end{figure}

Given their symmetry properties, pinacoids may potentially adopt the same orientation upon aligning one or more of their inertia axes, thus confering orientational order to the packing. To detect such an orientational order, the nematic order parameter $Q_{00}^2$ and the biaxial parameter $Q_{22}^2$ are computed in the packing bulk (see ref.~\cite{camenen_12,camp_97} for computation details). $Q_{00}^2$ assesses the highest level of alignement of a given inertia axis between all particles, either $\roarrow u$, $\roarrow v$, or $\roarrow w$ (see fig.~\ref{pinacoid}), whereas $Q_{22}^2$ assesses overall alignment of particle inertia axes as a consequence of the pinacoid symmetry properties. Figure~\ref{fig:Qoo} depicts the variations of $Q_{00}^2$ for pinacoid packings incorporating various proportions by volume of small, elongated, flat or flat \& elongated particles. It is remarkable that substituting small isometric pinacoids for large ones decreases the packing nematic order parameter whatever the substituted proportion. Similarly, substituting elongated particles for large isometric ones does not significantly increase the packing nematic order parameter. By contrast, substituting more than $30\%$ of flat or flat \& elongated particles for large isometric ones increases the packing nematic order parameter beyond the isotropic-nematic transition~\cite{allen_93,camp_97}, since flat particles tend to align their $\roarrow w$ inertia axis with the $\roarrow z$ direction as shown on fig.~\ref{fig:rendus2}. This alignment evidences the reorientation of flat and flat \& elongated pinacoids during the densification phase to form a layered structure which minimizes the potential energy of the equilibrated packing. This reorientation is facilitated by frustration release resulting from one particle dimension being significantly smaller than the other two. Besides, with the biaxial parameter never exceeding $0.05$ whatever the tested mixture, it should be pointed out that none of the tested mixture has reached the nematic-biaxial transition~\cite{camp_97}, in other words particle orientations remained random in the horizontal plane.

\subsubsection{Contact network in the samples}

In this subsection, we first check whether the contact network of frictionless isometric pinacoid packings fulfills the so-called isostatic conjecture. Next, we investigate whether gradually substituting small, elongated, flat, or flat \& elongated pinacoids for isometric ones modifies the typology of the contact network.

The isostatic conjecture states that amorphous rigid packings of hard frictionless particles are isostatic~\cite{alexander_98}. In this statement, \emph{rigid packing} refers to particle assembly which cannot be deformed subject to arbitrarily low stress without breaking any interparticle contact or deforming any particle, in other words packing in a mechanically stable equilibrium state. Furthermore, \emph{isostatic} means that the total number of interparticle constraints equals the sum of particle degrees of freedom (DOF), that is the coordination number equals twice the particle DOF. Though disputed in the recent years~\cite{smith_10}, authors have established the validity of this conjecture for random close packings of spheres~\cite{silbert_02,agnolin_07} and nontiling platonic solids such as tetrahedra, octaedra, icosaedra and dodecaedra~\cite{jaoshvili_10,jiao_11}, whereas packings of rounded particle shapes such as moderately oblate or prolate ellipsoids were found hypostatic~\cite{donev_04}.

The coordination number calculated here for monodisperse isometric pinacoid packings is $8.4\pm0.2$. Note that this value lies between those observed for similar particle shapes, such as maximally randomly jammed monodisperse tetrahedra for which $N\approx8.5$ and $N=8.6\pm0.1$ are respectively reported in ref.~\cite{neudecker_13,smith_10}, or monodisperse octahedra for which $N\approx7.7$ and $N=7.8\pm0.1$ are respectively reported in ref.~\cite{smith_10,jiao_11}. However, this value is well below the isostatic number $N=12$ since it amalgamates diverse contact types corresponding to different numbers of constaints. Indeed, upon assigning $1$ constraint to each vertex-face or edge-edge contact (simple contact), $2$ constraints to each edge-face contact (double contact) and $3$ constraints to each face-face contact (triple contact)~\cite{jaoshvili_10,jiao_11,camenen_12}, the number of constraints writes $N_c = N_s + 2.N_d + 3.N_t$, where $N_s$, $N_d$ and $N_t$ are, respectively, the numbers of simple, double and triple contacts. With $N_s=5.57\pm0.06$, $N_d=2.12\pm0.04$ and $N_t=0.72\pm0.07$, we obtain $N_c=11.99\pm0.35$. Despite the finite size of our packing periodic cell, observe that this value compares well with the isostatic number $N=12$ valid for infinite size packings. As a consequence, our monodisperse isometric pinacoid packings are reasonably isostatic as conjectured in reference~\cite{alexander_98}. Yet, it should be observed that the number of face-face contacts per particle ($N_t=0.72\pm0.07$) is lower than those tabled in ref.~\cite{jiao_11} for tetrahedra ($2.21\pm0.01$), even for octaedra ($1.44\pm0.01$). As observed in ref~\cite{neudecker_13}, we argue that the probability of perfect face-face alignment is low compared to that of either slightly shifted face-face or low angle edge-face contact. This is consistent with the finite slope/moderate first peak value of our pair correlation curves (see Fig.~\ref{fig:gr}) and our significantly higher number of edge-face contacts per particle ($N_d=2.12\pm0.04$) compared to those reported for tetrahedra ($0.98\pm0.01$) and octaedra ($1.38\pm0.01$) in reference~\cite{jiao_11}.

\begin{figure*}[!t]
	\centering
\begin{tabular}{c c}
\includegraphics*[width=0.85\columnwidth]{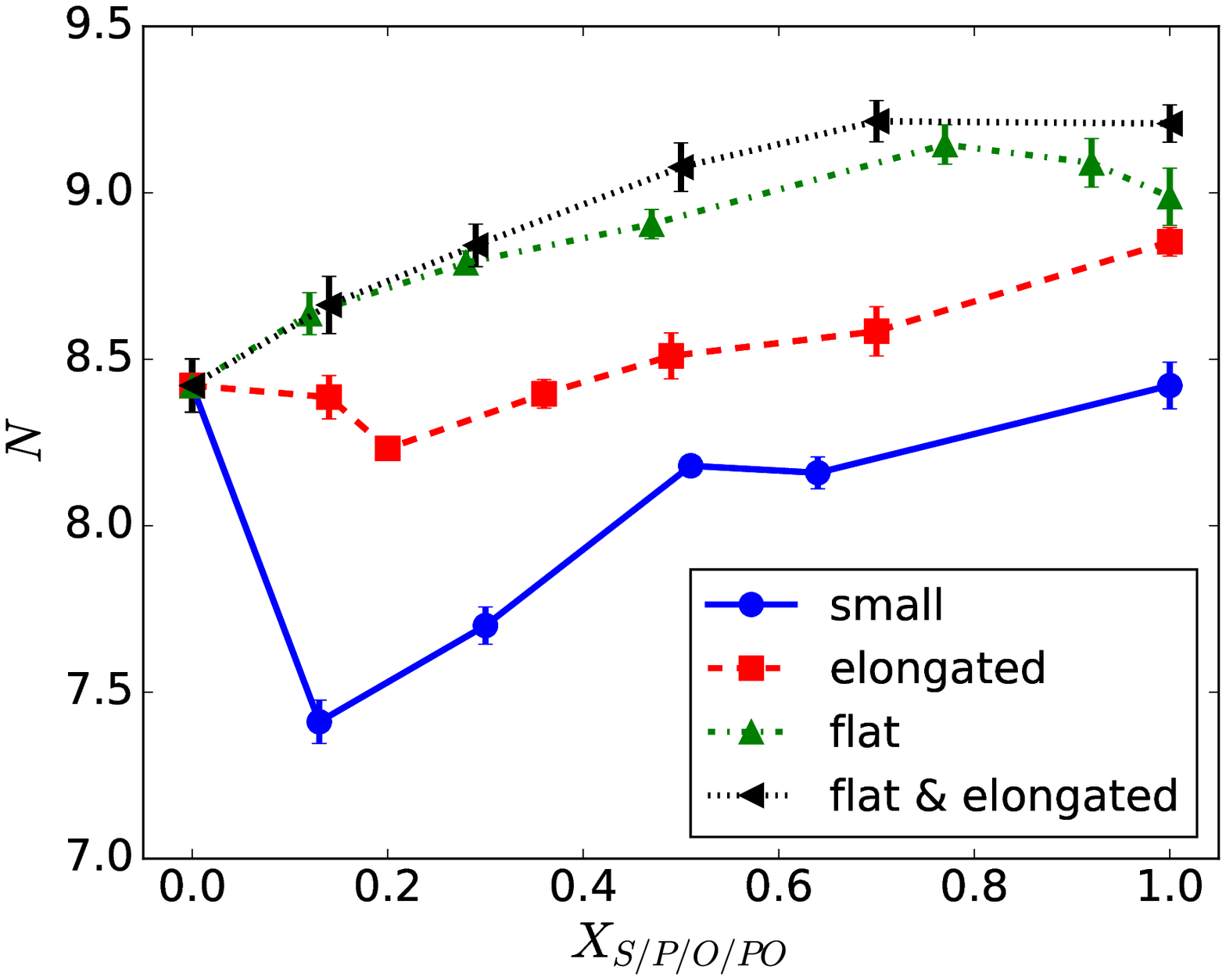}&\includegraphics*[width=0.85\columnwidth]{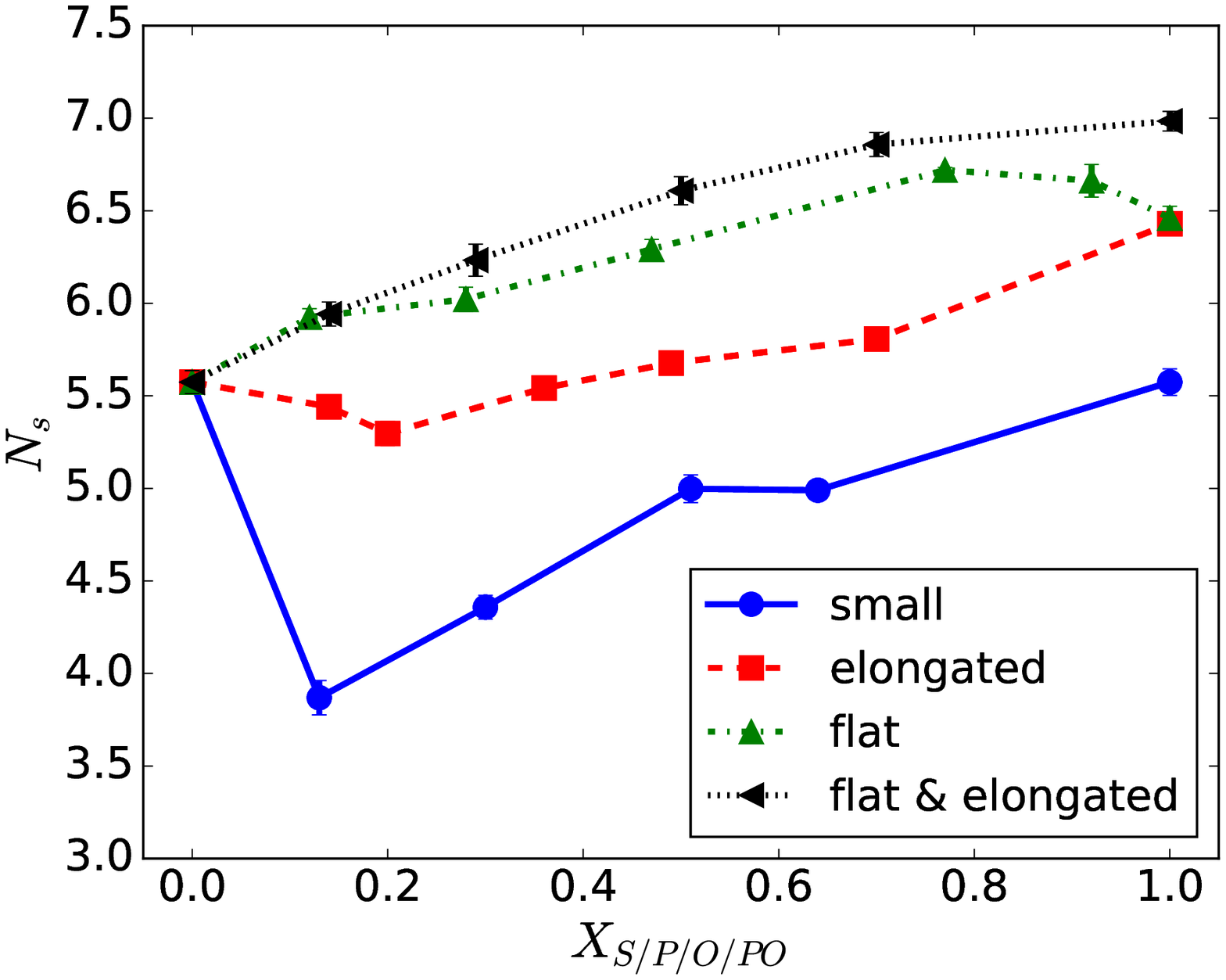}\\
(a) & (b)\\
\includegraphics*[width=0.85\columnwidth]{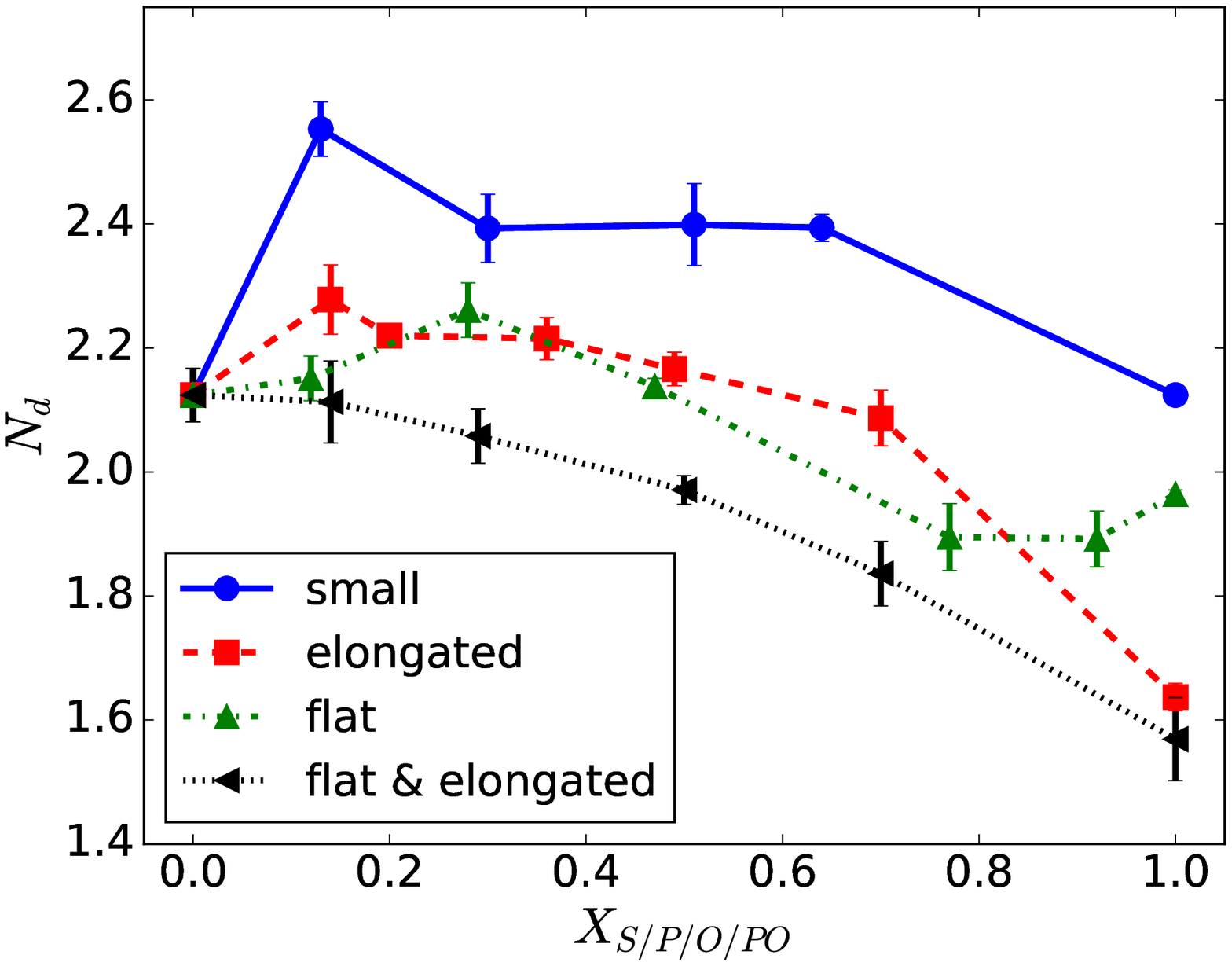}&\includegraphics*[width=0.85\columnwidth]{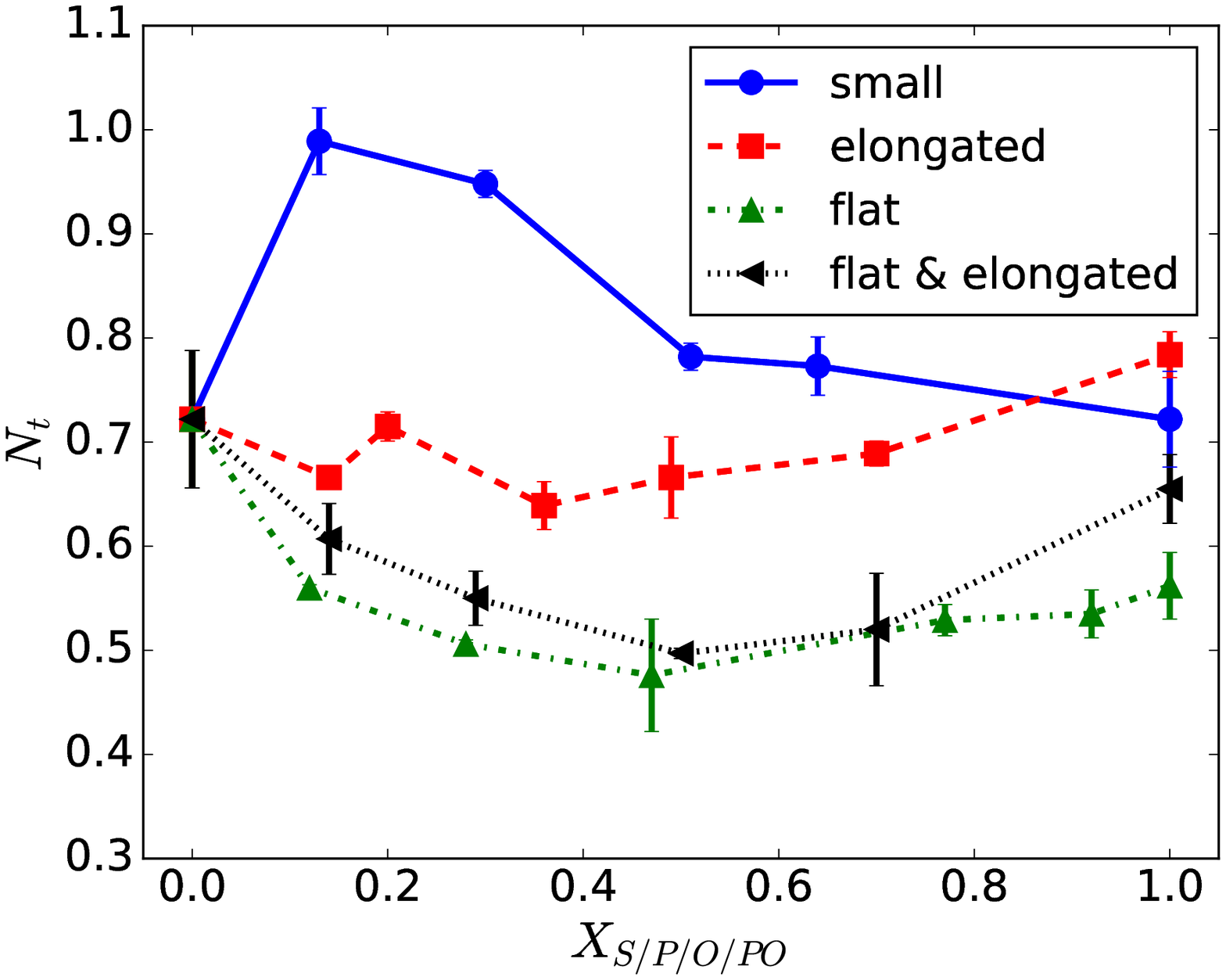}\\
(c) & (d)
\end{tabular}
\caption{\label{fig:coor}(Color online) Coordination number $N$ of $(a)$ all, $(b)$ simple, $(c)$ double, and $(d)$ triple contacts as a function of the proportion by volume of small ($X_S$), elongated ($X_P$), flat ($X_O$) or flat \& elongated ($X_{PO}$) pinacoids in the packing. Error bars denote the standard deviation}
\end{figure*}

To get insight into how the contact network fluctuates with increasing proportions of small or non-isometric particles, it is worth examining Fig.~\ref{fig:coor} depicting the variations of the coordination numbers of all, simple, double and triple contacts as a function of the proportion by volume of small, elongated, flat, or flat \& elongated pinacoids. When gradually substituting small isometric pinacoids for large ones, the coordination number first reaches a minimum $N=7.4\pm0.2$ for $X_S=13\%$ and then gradually increases back to its monodisperse value, as shown on Fig.~\ref{fig:coor}a. Interestingly, similar coordination number fluctuations may be observed with simple contacts (Fig.~\ref{fig:coor}b), whereas the coordination numbers of double and triple contacts vary conversely (Fig.~\ref{fig:coor}c and~\ref{fig:coor}d). In fact, small pinacoids are trapped inside the excluded volume of large ones. Until their proportion is sufficient for completely filling this volume, which corresponds to the achievement of the packing maximum solid fraction ($X_S=30\%$, see Fig.~\ref{fig:phi_particles}b), their low steric hindrance allows them to rotate and establish more stable face-face or edge-face contacts with large ones. Contrarywise, gradually substituting elongated, flat, or flat \& elongated pinacoids for isometric ones tends to increase the coordination number $N$ of our pinacoid packings up to, respectively, $8.9\pm0.1$, $9.1\pm0.2$, and $9.2\pm0.2$. Actually, these non-isometric particles are too large to fit in the excluded volume of large isometric pinacoids while leaving there arrangement undisturbed. Besides, contrary to elongated or isometric pinacoids, the adjacent trapezoidal flat faces of flat and flat \& elongated pinacoids do not intersect at right angle and these particles have a smaller thickness than the former. Stated otherwise, flat and flat \& elongated pinacoids have a broken angular symmetry and reduced external surface (at least reduced triangular flat faces for flat \& elongated pinacoids), causing the coordination number of face-face contacts to decrease to the benefit of simple contacts with increasing proportions of flat or flat \& elongated pinacoids in the packing (see Fig.~\ref{fig:coor}b and d). Furthermore, for increasing proportions of flat or flat \& elongated pinacoids, observe that the coordination number of simple contacts increases quicker than the coordination number of face-face contacts decreases, which is a consequence of packing isostaticity : whatever the proportion of small, elongated, flat, or flat \& elongated particles, all our packings remain isostatic with $12.02\pm0.35$ constraints per particle.

\begin{figure}[!t]
	\centering
\includegraphics*[width=0.9\columnwidth]{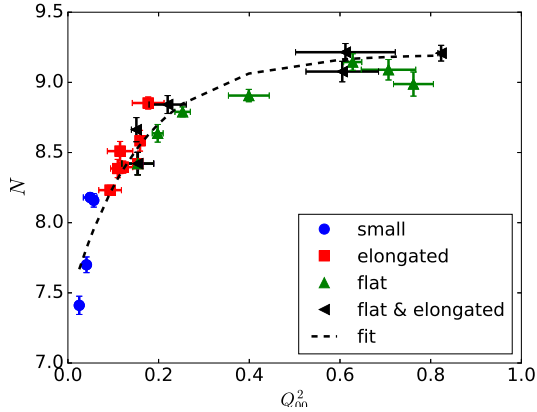}
\caption{\label{fig:N_Qoo}(Color online) Coordination number $N$ $vs$ the nematic order parameter $(Q_{00}^2)$ of pinacoid packings incorporating various proportions by volume of small ($X_S$), elongated ($X_P$), flat ($X_O$) or flat \& elongated ($X_{PO}$) pinacoids. Fit equation is $N(Q^2_{00}) = 7.4 + (9.2 - 7.4).[1 - \exp(-Q^2_{00}/0.2)]$ with $R^2=0.92$. Error bars denote the standard deviation.}
\end{figure}

Finally, Fig.~\ref{fig:N_Qoo} depicts the variations of the coordination number as a function of the nematic order parameter. This figure suggests that the coordination number increases continuously with orientational ordering in the packings and irrespective of particle size or aspect ratio. An exponential fit of the following form was adjusted to the point cloud:

\be N(Q^2_{00}) = N(0)+[N(1)-N(0)].[1-\exp(-\frac{Q^2_{00}}{Q^2_{00,c}})] \ee \label{fit}

\noindent where [$N(0),N(1),Q^2_{00,c}$] were calculated respectively as ($7.4, 9.2,0.2$) with $R^2=0.92$. This fit suggests that orientationnally ordered pinacoid packings have coordination numbers in excess of about $9$ and, conversely, that packings may be considered orientationnally disordered for coordination numbers below roughly $8.5$, corresponding to $Q^2_{00}\leq 0.2$.

\section{Conclusion}\label{sec:conclusion}

The properties of dense packings of spheres or pinacoids compacted under their own weight have been investigated using three-dimensional Non Smooth Contact Dynamic simulations. Various proportions by volume of small, elongated, flat, or flat \& elongated pinacoids were substituted for large isometric ones in order to understand how polydispersity and shape affect their solid fraction and microstructural properties. Numerical simulations show that disordered assemblies of frictionless pinacoids, were they monodisperse or bidisperse, pack with a higher solid fraction than corresponding assemblies of spherical or rounded particles, thus fulfilling the analogue of Ulam's conjecture for random packings proposed in ref.~\cite{jiao_11}. This seeming discrepancy with experimental results reported in ref.~\cite{delarrard_99,baker_10} is believed to lie with difficulties in overcoming interparticle friction through experimental densification processes. Moreover, solid fraction increases further with bidisperse particles and peaks when the proportion of small ones reaches $30\%$, achieving $\phi=0.726\pm0.004$ and $\phi=0.769\pm0.001$ respectively for spheres and pinacoids. Contrarywise, partial substitution of flat pinacoids for isometric ones results in packing solid fraction decrease by a maximum $8\%$, especially when flat particles are also elongated. Minimum solid fraction is achieved when the proportion by volume of flat or flat \& elongated pinacoids reaches $50\%$. Nevertheless, particle shape seems to play a minor role on packing solid fraction compared to polydispersity. Additional investigations focused on packing microstructure confirm that, with $12.02\pm0.35$ constraints per particle, pinacoid packings fulfill the isostatic conjecture and that they are free of order except beyond $30$ to $50\%$ by volume of flat or flat \& elongated polyhedra in the packing. This order increase progressively takes the form of a nematic phase as flat or flat \& elongated particles reorientate so that their largest projected area is horizontal to minimize the packing potential energy. Simultaneously, this reorientation seems to increase the solid fraction value slightly above the maximum achieved by monodisperse isometric pinacoids, as well as the coordination number. Finally, partial substitution of elongated pinacoids for isometric ones has limited effect on packing solid fraction or order.


\begin{acknowledgments}
The authors thank the team running the Centre de Calculs Intensifs des Pays de la Loire (CCIPL) for providing the calculation ressources through the MTEEGD project. They are also grateful to their colleagues J.N. Roux, N. Roquet and P. Richard for helpful conversations. Many thanks to the LMGC90 team in Montpellier for making their software plateform freely available.
\vspace{4cm}
\end{acknowledgments}



\bibliographystyle{apsrev}

\end{document}